\documentclass[a4paper,10pt]{article}
\usepackage{graphicx}
\usepackage{comment}
\usepackage{amssymb}
\usepackage{amsmath}
\usepackage{nicefrac}
\usepackage{graphicx}
\usepackage{dcolumn}
\usepackage{bm}
\usepackage{comment}
\usepackage{multirow}
\usepackage{cite}
\usepackage{mathrsfs}

\makeatletter
\@ifpackageloaded{MnSymbol}{}{%
\DeclareFontFamily{OMX}{MnSymbolE}{}
\DeclareSymbolFont{largesymbolsMn}{OMX}{MnSymbolE}{m}{n}
\SetSymbolFont{largesymbolsMn}{bold}{OMX}{MnSymbolE}{b}{n}
\DeclareFontShape{OMX}{MnSymbolE}{m}{n}{
<-6> MnSymbolE5
<6-7> MnSymbolE6
<7-8> MnSymbolE7
<8-9> MnSymbolE8
<9-10> MnSymbolE9
<10-12> MnSymbolE10
<12-> MnSymbolE12}{}
\DeclareFontShape{OMX}{MnSymbolE}{b}{n}{
<-6> MnSymbolE-Bold5
<6-7> MnSymbolE-Bold6
<7-8> MnSymbolE-Bold7
<8-9> MnSymbolE-Bold8
<9-10> MnSymbolE-Bold9
<10-12> MnSymbolE-Bold10
<12-> MnSymbolE-Bold12}{}
\DeclareMathDelimiter{\llangle}{\mathopen}{largesymbolsMn}{'164}{largesymbolsMn}{'164}
\DeclareMathDelimiter{\rrangle}{\mathclose}{largesymbolsMn}{'171}{largesymbolsMn}{'171}
}
\makeatother

\newcommand*{\Eulerian}[2]{%
\mathinner{%
\genfrac\llangle\rrangle{0pt}{}{#1}{#2}%
}%
}

\begin{document}

\huge

\begin{center}
Schr\"odinger equation on a generic radial grid
\end{center}

\vspace{0.5cm}

\large

\begin{center}
Christopher Bowen$^a$ and Jean-Christophe Pain$^{a,b,}$\footnote{jean-christophe.pain@cea.fr}
\end{center}

\normalsize

\begin{center}
\it $^a$CEA, DAM, DIF, F-91297 Arpajon, France\\
\it $^b$Universit\'e Paris-Saclay, CEA, Laboratoire Mati\`ere en Conditions Extr\^emes,\\
\it 91680 Bruy\`eres-le-Ch\^atel, France
\end{center}

\begin{abstract}
In this note, we discuss the choice of radial grid in the numerical resolution of the Schr\"odinger equation. We detail the transformation of the equation resulting from a change of variable and function for a generic radial grid, using either the explicit or implicit form of the relation describing the change of variable, and apply it to the $a~r+b~\ln(r)$ log-linear mesh. It is shown that, in the former case, the first three derivatives of the Lambert $W$ function are required. This complication becomes unnecessary if we adopt the implicit relation instead.
\end{abstract}

\section{Introduction}

We wish to solve a general Schr\"odinger-type equation
\begin{equation}
P''(r)-q(r)P(r)=\kappa(r).
\label{eq:Numerov1}
\end{equation}
This form arises from using atomic units and writing the radial part of the wavefunction in the form $P(r)=rR(r)$, with normalisation condition $\int_0^{\infty} P^2(r) dr = 1$. $q(r)$ has the form
\begin{equation*}
q(r) = \frac{\ell(\ell+1)}{r^2}+ V(r) - E,
\label{eq:Numerov2}
\end{equation*}
where $\ell$ is the orbital quantum number, $V(r)$ the electronic potential, $E$ the energy and the non-homogeneous term $\kappa(r)$ typically arises from a Hartree-Fock treatment of the electronic structure \cite{Hartree1957}. We know how to solve such an equation on a uniform radial grid (using for instance Numerov's method \cite{Numerov1933,Cooley1961,Hajj1974,Nikiforov2005}). However, in that case, the error near the origin becomes as large as the wavefunction itself for $\ell\geq 10$ \cite{Havlova1983}. Therefore, realistic atomic physics calculations require grids that better resolve the wavefunction in the region near the nucleus where the potential varies quickly, but with larger steps elsewhere. Two such grids are the exponential $r=e^x$ and log-linear $x=a r + b \ln{}(r)$, where $x$ is a linear (constant step) parameter. The question that we want to address is how to transform the radial Schr\"odinger equation such that, on the uniform grid, it keeps its initial form (\ref{eq:Numerov1}). We first note that in the case of the exponential grid, the non-uniform variable $r$ appears as an explicit function of the linear parameter, whereas in the second case, it is written in implicit form. As discussed in section \ref{sec:hybrid}, $r$ may be written explicitly in terms of the Lambert $W$ function, but we will see that this complicates the algebra. Even though the log-linear grid is well known, we have not found in the literature a discussion of the equation that results from the implicit change of variables, and thus expect that the analysis below should prove useful in general. 

In section \ref{sec:transfo}, we show how to transform the Schr\"odinger equation using the explicit and implicit forms of the change of variables, and check their equivalence. In section \ref{sec:hybrid}, we apply the formalism to the hybrid (log-linear) grid and compare the amount of algebra involved, thereby illustrating the necessity of using the most convenient approach. 

\section{Transformation of the Schr\"odinger equation}\label{sec:transfo}

\subsection{Changes of variable and function}\label{subsec:changes}

Let us start with the explicit case. We write $r = f(x)$ and look for a new function $Y(x)$ that obeys $Y''(x)-p(x)Y(x)=h(x)$. The solution is given in \cite{Havlova1983} by Havlov\'a and Smr$\mathrm{\check{c}}$ka for the homogeneous case corresponding to $\kappa(r)=0$ in Eq. (\ref{eq:Numerov1}). The authors factorize $P$ in the form $P(r) =  g(x)Y(x)$,
and simply state the result for $g$ such that the wave equation for $Y$ has the required form: $g^2(x)=f'(x)$ and
\begin{equation}
p(x) = q(f(x)) (f'(x))^2 - \frac{f'(x)}{g(x)}\left(\frac{g'(x)}{f'(x)}\right)'.
\label{eq:havlova}
\end{equation}
The demonstration is straightforward, we give it here since it gives us the opportunity to include the inhomogeneous term. Throughout the paper, the prime symbol represents the first derivative, double prime the second derivative, etc. It should be clear from the context with respect to what variable we are differentiating. We will often omit, for simplicity and when there is no ambiguity, the explicit dependence of the function with respect to its variable. One has, with $x$ as the independent variable, 
\begin{equation*}
\frac{dP}{dr}=\frac{dP}{dx}\frac{dx}{dr}=\frac{g'Y+gY'}{f'}
\end{equation*}
and
\begin{equation*}
\frac{d^2P}{dr^2}=\frac{d}{dx}\left(\frac{dP}{dr}\right)\frac{dx}{dr}=\frac{(g''Y+2g'Y'+gY'')f'-f''(g'Y+gY')}{f'^3}.
\end{equation*}
Therefore, (\ref{eq:Numerov1}) becomes
\begin{equation*}
\frac{d^2P}{dr^2}-qP=\frac{(g''Y+2g'Y'+gY'')f'-f''(g'Y+gY')-qf'^3gY}{f'^3}=\kappa,
\end{equation*}
yielding
\begin{equation}\label{eq:int}
Y''(gf')+Y'(2g'f'-f''g)+Y(g''f'-f''g'-qf'^3g)=\kappa f'^3.
\end{equation}
We now find $g$ by requiring that the term in $Y'$ vanish,
\begin{equation}
2g'f' = f'' g \Rightarrow 2 \ln{g} = \ln{f'} ,\;\;\; g^2 = Af'.
\end{equation}
We can set the integration constant $A=1$, since any scalar multiple of $g$ will factor out in the homogeneous case, or can be absorbed into the overall normalisation of the wavefunction when $\kappa \neq 0$. Eq. (\ref{eq:int}) then becomes
\begin{equation*}
Y''+Y\frac{(g''f'-f''g'-qf'^3g)}{gf'}=\frac{\kappa f'^2}{g}
\end{equation*}
leading, since $Y''-pY=h$ to
\begin{equation*}
p=qf'^2-\frac{f'}{g}\left(\frac{g''f'-f''g'}{f'^2}\right)=qf'^2-\frac{f'}{g}\left(\frac{g'}{f'}\right)'
\end{equation*}
and
\begin{equation*}
h=\frac{\kappa f'^2}{g} = \kappa (f')^{3/2}.
\end{equation*}
Havlov\'a and Smr$\mathrm{\check{c}}$ka suggested that the transformation $r=f(x)=x^3$ represents a reasonable compromise between the linear and logarithmic meshes.
In that case, $f'(x)=3 x^2$, $h(x)=3 \sqrt{3} x^3 \kappa(r)$ and 
\begin{equation*}
p(x)=9x^4 q(r) + \frac{2}{x^2}= 9x^4 (V(r)-E)+ \frac{(3\ell+2)(3\ell+1)}{x^2}.
\end{equation*}
We can also easily find the well-known result for the exponential grid, $f(x)=e^x$.
Then 
\begin{equation*}
g(x)=e^{x/2}, \;\;\; \frac{f'(x)}{g(x)}\left(\frac{g'(x)}{f'(x)}\right)'=-1/4
\end{equation*}
and
\begin{equation*}
p(x)= \left[V(r)-E + \frac{\ell(\ell+1)}{r^2} \right] e^{2x} + \frac{1}{4}
= r^2(V(r)-E) + \left(\ell+\frac{1}{2}\right)^2 .
\end{equation*}
Now let us look at the implicit form, writing $Y(x)=P(r(x))$. The first two derivatives of $P$ with respect to $r$ are respectively
\begin{equation*}
P'(r) = P'(r(x)) \frac{dx}{dr} = Y'(x) \frac{dx}{dr}
\end{equation*}
and
\begin{equation*}
P''(r) = Y''(x) \left(\frac{dx}{dr}\right)^2+Y'(x)\frac{d^2x}{dr^2}.
\end{equation*}
The differential equation becomes, in terms of the variable $x$:
\begin{equation*}
Y''(x)+\frac{d^2x}{dr^2}\left(\frac{dr}{dx}\right)^2Y'(x)-\left(\frac{dr}{dx}\right)^2 q(r(x))~Y(x)
=\left(\frac{dr}{dx}\right)^2 \kappa(r(x)),
\end{equation*}
i.e.
\begin{equation*}
Y''(x)+\alpha(x)~Y'(x)-\beta^2(x)~q(r(x))~Y(x)=\beta^2(x)~\kappa(r(x))
\end{equation*}
with
\begin{equation*}
\beta(x)\equiv \frac{dr}{dx}
\end{equation*}
and 
\begin{equation*}
\alpha(x)\equiv \left.\frac{d^2x}{dr^2}\right|_x \beta^2(x).
\end{equation*}
We now look for a change of function cancelling the first-order derivative term, in order to recover the initial form (\ref{eq:Numerov1}). We try to factorize $Y$ in the form
\begin{equation*}
Y(x)= t(x) s(x),
\end{equation*}
where $s$ is a function to be determined, $t$ being the new function after transformation. Using
\begin{equation*}
Y'=t's+ts', \;\;\;Y'' = t''s+2t's'+ts'',
\end{equation*}
the $t$ function satisfies
\begin{equation*}
t''s+2t's'+ts''+\alpha t's+\alpha ts'-\beta^2qts=\beta^2 \kappa
\end{equation*}
or
\begin{equation*}
t''+\left(\frac{2s'}{s}+\alpha\right)t'+\left(\frac{s''+\alpha s'}{s}-\beta^2q\right)t=\frac{\beta^2 \kappa}{s}.
\end{equation*}
The function $s$ is chosen in order to cancel the first-order term, thus
\begin{equation*}
\frac{s'}{s}=-\frac{\alpha}{2} 
\end{equation*}
whose solution is
\begin{equation*}
s(x)=\exp\left[-\frac{1}{2}\int dx'\alpha(x')\right].
\end{equation*}
We calculate the primitive by reverting to the variable $r$:
\begin{equation*}
\int dx'\alpha(x')=\int dx'\left.\frac{d^2x}{dr^2}\right|_{x'}\left(\left.\frac{dx}{dr}\right|_{x'}\right)^{-2}=\int dr'\left. \frac{d^2x}{dr^2}\right|_{r'}\left(\left.\frac{dx}{dr}\right|_{r'}\right)^{-1}
=\ln\left(\frac{dx}{dr}\right),
\end{equation*}
yielding
\begin{equation*}
s(x)=\exp\left[\frac{1}{2}\ln\left(\frac{dr}{dx}\right)\right]=\sqrt{\beta(x)}.
\end{equation*}
The coefficient of the zero-order term reads, differentiating the above condition with respect to $x$:
\begin{equation*}
\frac{s''s-s'^2}{s^2}=\frac{s''}{s}-\left(\frac{s'}{s}\right)^2=-\frac{\alpha'}{2}
\end{equation*}
where $\alpha'(x)=d\alpha/dx$. One finds
\begin{equation*}
\frac{s''}{s}=\frac{\alpha^2}{4}-\frac{\alpha'}{2}
\end{equation*}
and
\begin{equation*}
\frac{s''}{s}+\alpha\frac{s'}{s}=-\frac{\alpha^2}{4}-\frac{\alpha'}{2}.
\end{equation*}
Finally, the differential equation in $t$ is given by the expression
\begin{equation}
t''-\left(\beta^2 q+\frac{\alpha^2}{4}+\frac{\alpha'}{2}\right)t=\frac{\beta^2\kappa}{s}=\beta^{3/2} \kappa
\label{eq:paindejonghe}
\end{equation}
and is in the desired form
\begin{equation*}
t''(x)-p(x)t(x)=h(x)
\end{equation*}
with
\begin{equation}\label{eq:hatf}
p=\beta^2 q+\frac{\alpha^2}{4}+\frac{\alpha'}{2}
\end{equation}
and 
\begin{equation}\label{eq:hatk}
h=\beta^ {3/2} \kappa.
\end{equation}
To summarize, to every change of variables $r\rightarrow x(r)$, we associate the change of function $P(r)\rightarrow t(x)$, with 
\begin{equation*}
P(r(x)) = Y(x)= t(x)/\sqrt{\frac{dx}{dr}}.
\end{equation*}
The function $t$ then satisfies a differential equation in terms of the variable $x$, solvable by the Numerov method, the functions $p$ and $h$ being given by the above relations (\ref{eq:hatf}) and (\ref{eq:hatk}). For instance, in the case of the cubic scale, one has
\begin{equation*}
\beta(x)=3x^2, \;\;\;\; \alpha(x)=-\frac{2}{x}
\end{equation*}
and
\begin{equation*}
t''-\left(9x^4 q +\frac{2}{x^2}\right)t=3\sqrt{3}x^3 \kappa
\end{equation*}
as before. 

\subsection{Equivalence}\label{subsec:equi}

It is straightforward to show that the two forms (\ref{eq:paindejonghe}) and (\ref{eq:havlova}) are equivalent. Since we have defined $\beta = f'(x)$, the first terms are equal, and we just need to show
\begin{equation}
\frac{\alpha^2}{4}+\frac{\alpha'}{2} = - \frac{f'}{g}\left(\frac{g'}{f'}\right)'.
\label{eq:equiv}
\end{equation}
The right-hand side of Eq. (\ref{eq:equiv}) is equal to 
\begin{eqnarray*}
- \frac{f'}{g}\left(\frac{g'}{f'}\right)'
&=&
- \sqrt{f'} \left(\frac{f''}{2 (f')^{3/2}}\right)' 
= \frac{-2 (f')^2 f''' + 3 f' (f'')^2}{4 (f')^3} \\
&=& -\frac{f'''}{2 \beta} + \frac{3(f'')^2}{4 \beta^2}.
\end{eqnarray*}
Now
\begin{equation*}
\frac{d^2x}{dr^2} = \frac{d}{dr}\left(\frac{1}{\beta}\right) = - \frac{f''}{\beta^3}
\end{equation*}
and using the definition of $\alpha$, we find
\begin{equation*}
f'' = - \beta^3 . \frac{\alpha}{\beta^2} = - \alpha \beta.
\end{equation*}
Thus
\begin{equation*}
f''' = - \beta \alpha' - \alpha \beta' = - \beta \alpha' + \alpha^2 \beta
\end{equation*}
and 
\begin{equation*}
- \frac{f'''}{2 \beta} + \frac{3(f''){^2}}{4 \beta^2} = 
\frac{\beta \alpha' - \alpha^2 \beta}{2 \beta} + \frac{3(-\alpha \beta)^2}{4 \beta^2}
= \frac{\alpha'}{2}+ \frac{\alpha^2}{4},
\end{equation*}
as required. 

\section{Hybrid radial grid}\label{sec:hybrid}

\subsection{The $a~r+b~\ln r$ mesh}\label{subsec:arblnr}

In solving the equation numerically, it is convenient to resort to a mesh of equidistant points in the independent variable. However, since the potential $V(r)$ has a pole at $r=0$, the number of points needs to be very dense in the vicinity of $r=0$, while there is no need to take a large number of points far from the origin. One possibility to handle that problem is to replace the radial variable $r$ by a new variable obtained by a suitable transformation, ensuring that an equidistant mesh in the new variable corresponds to a mesh whose step size decreases as $r$ approaches the origin. In that framework, an interesting possibility consists in using the transformation (see for instance Refs. \cite{Bratcev1965,Froese1977,Pain2006}):
\begin{equation}\label{eq:mixte}
x=a~r+b~\ln r
\end{equation}
where $a$ and $b$ are constants. Such a transformation is more convenient than the procedure proposed by Herman and Skillman \cite{Herman1963}, consisting in dividing the integration mesh into several blocks and doubling the mesh interval in each subsequent block for large $r$. The mesh (\ref{eq:mixte}) is well adapted to excited-state wavefunctions as $x\approx a~ r$ when $r\rightarrow\infty$. In practice, Chernysheva \emph{et al.} suggested taking $x_{\mathrm{min}}\leq x\leq x_{\mathrm{max}}$ where
\begin{equation*}
x_{\mathrm{min}}=-b~(10+\ln Z),
\end{equation*}
and $Z$ represents the atomic number \cite{Chernysheva1976,Chernysheva1999}. Such a lower bound corresponds to $r_{\mathrm{min}}\approx4.5~ 10^{-5}/Z$. 
The mesh of equidistant points in $x$ is $x_i=x_0+i~ h$, with $i=0, 1, \cdots, N$, and thus $x_0=x_{\mathrm{min}}$ and $x_N=x_{\mathrm{max}}$. The coefficient $a$ can be determined by
\begin{equation*}
a=\frac{x_{\mathrm{max}}-b\ln r_{\mathrm{max}}}{r_{\mathrm{max}}},
\end{equation*}
while the coefficient $b$, step size $h$, number of interval points $N$, and $r_{\mathrm{max}}$ (maximum value of $r$) are input data \cite{Chernysheva1976,Chernysheva1999}. Alternatively, one can input $r_{\mathrm{min}}$, $r_{\mathrm{max}}$, $a$, $b$ and $N$, and deduce the step size 
 \cite{Wilson2011,Pain2011}.

The change of variables is in that case $x(r)=ar+b~\ln(r)$, $r^be^{ar}=e^x$,
\begin{equation*}
\frac{dx}{dr}=\frac{1}{\beta}=\frac{b}{r}+a, \;\;\;\;\frac{d^2x}{dr^2}=-\frac{b}{r^2}
\end{equation*}
giving the simple forms
\begin{equation*}
\beta(x)=\frac{r}{b+ar}, \;\;\;\;\alpha(x)=-\frac{b}{(b+ar)^2}.
\end{equation*}
This yields
\begin{equation*}
\alpha'(x)=\frac{d\alpha}{dx}=\frac{2ab}{(b+ar)^3}\frac{dr}{dx}=\frac{2abr}{(b+ar)^4}
\end{equation*}
and
\begin{equation*}
s(x)=\sqrt{\beta(x)}=\sqrt{\frac{r}{b+ar}}.
\end{equation*}
Therefore, the differential equation for $t$ is
\begin{equation}
t''-\frac{1}{(b+ar)^2}\left[r^2 q+b\frac{b+4ar}{4(b+ar)^2}\right]t=
\left(\frac{r}{b+ar}\right)^{3/2} \kappa
\label{eq:loglin}
\end{equation}
with $r \equiv r(x)$ and
\begin{equation}\label{eq:eqy}
t(x)=Y(x)~\sqrt{\frac{b}{r}+a}.
\end{equation}
Of course, one recovers the relations for the linear grid when $a=1$, $b=0$:
\begin{equation*}
x(r)=r,\;\;\;\;\beta=1,\;\;\;\;t=Y, \;\;\;\;p=q, \;\;\;\;h=\kappa
\end{equation*}
and those for the logarithmic grid when $a=0$, $b=1$:

\begin{equation*}
x(r)=\ln(r), \;\;\;\;\beta=r, \;\;\;\;t(x)=\frac{Y(x)}{\sqrt{r}}
\end{equation*}
and 
\begin{equation*}
t''-\left(r^2 q+\frac{1}{4}\right)t=r^{3/2} \kappa. 
\end{equation*}

\subsection{Explicit inversion for the log-linear case}\label{subsec:inversion}

One has $x=ar+b\ln r$ and therefore the inversion formula 
\begin{equation*}
r=\exp\left[\frac{x-ar}{b}\right],
\end{equation*}
which can be put in the form
\begin{equation*}
\frac{a}{b}r~\exp\left[\frac{ar}{b}\right]=\frac{a}{b}\exp\left[\frac{x}{b}\right]
\end{equation*}
yielding
\begin{equation}\label{eq:reci}
r=\frac{b}{a}~W\left(\frac{a}{b}\exp\left[\frac{x}{b}\right]\right),
\end{equation}
where $W$ is the Lambert function \cite{Wilson2011,Pain2011,Corless1996} satisfying $y~ e^y=x \Rightarrow y=W(x)$.

The derivation showing the formal equivalence of the explicit and implicit formulations is indicative that in terms of algebra the two approaches are not equivalent. Using the Havlov\'a-Smr$\mathrm{\check{c}}$ka form is straightforward for the simple explicit relations $r(x)$ above, but becomes cumbersome for the log-linear grid, involving third derivatives of the Lambert $W$ function. The algebra involved in obtaining the differential equation (\ref{eq:loglin}) can become rapidly tedious depending on which starting point is adopted (see \ref{appA}). 
It is easiest to start from the form
\begin{equation*}
W'(x)=\frac{W(x)}{x\left[1+W(x)\right]}.
\end{equation*}
Let $u=(a/b ) \exp{(x/b)}, \;\;\; u_x=du/dx=u/b$. Then with $r=f(x)=(b/a) \; W(u)$, we find
\begin{equation*}
f'(x)=\frac{b}{a} W'(u) u_x = \frac{b}{a} \frac{W(u)}{u\left[1+W(u)\right]} \frac{u}{b} = \frac{1}{a} \frac{W(u)}{\left[1+W(u)\right]} 
\end{equation*}

\begin{equation*}
f''(x) = \frac{1}{a} \frac{\left[1+W(u)\right]W'(u) - W(u)W'(u)}{\left[1+W(u)\right]^2} u_x =\frac{1}{ab} \frac{W(u)}{\left[1+W(u)\right]^3}
\end{equation*}
and
\begin{equation*}
f'''(x) = \frac{1}{ab} \frac{\left[1+W(u)\right]^3 W'(u)- 3W(u) \left[1+W(u)\right]^2W'(u)}{\left[1+W(u)\right]^6} u_x 
        = \frac{1}{ab^2} \frac{W(u)-2W^2(u)}{\left[1+W(u)\right]^5}.
\end{equation*}
Hence 
\begin{equation*}\label{eq:run}
- \frac{f'''}{2 f'} + \frac{3(f''){^2}}{4 (f')^2}
=\frac{1+4W(u)}{4b^2(1+W(u))^4} = \frac{1+4 ar/b}{4b^2(1+ar/b)^4} = b \frac{b+4 ar}{4(b+ar)^4}.
\end{equation*}
Similarly,
\begin{equation*}\label{eq:rdeux}
(f')^2 = \frac{1}{a^2}\left[\frac{W(u)}{1+W(u)}\right]^2 = \frac{1}{a^2}\left[\frac{ar/b}{1+ar/b}\right]^2 = 
\frac{r^2}{(b+ar)^2}
\end{equation*}
and we recover the desired form (\ref{eq:loglin}), albeit with much more work.

It is worth mentioning that the above results can be generalized to the case of a mesh of the type $x=a~r^{k}+b~\ln r$. Indeed, setting $\tilde{r}=r^k$ yields $x=\tilde{a}~\tilde{r}+~\tilde{b} \ln \tilde{r}$ with $\tilde{a}=a$ and $\tilde{b}=b/k$. The formulas for the corresponding implicit and explicit transformations are given in \ref{appB}. Such a mesh combines for instance the $x=r^{1/3}$ form proposed by Havlov\'a and Smr$\mathrm{\check{c}}$ka, and the hydrid character introduced by the addition of the logarithmic term. However, we have not investigated 
its usefulness in practice.

\section{Conclusion}\label{sec:conclu}

We have presented the general form of the differential equation obtained after a general change of variables on a radial grid. The transformation relations resulting from the implicit change of variable have been shown as expected to be equivalent to those given by Havlov\'a and Smr$\mathrm{\check{c}}$ka, but are logically the easiest ones to use in the case of a log-linear grid, and presumably in every case where the implicit relation provides the simplest form. They will of course be the only ones usable when 
no explicit relation can be found. Note that aside from shooting methods such as the Numerov algorithm, other techniques exist, such as the Wronskian method \cite{Fernandez2011} or the canonical function method \cite{Tannous2008}. Irrespective of the choice of numerical method, 
it will be essential to express the transformed ordinary differential equation in the simplest possible form, in order to simplify coding and minimize computational cost as well as numerical errors. We hope that the discussion in this note will prove useful in that regard. 

\appendix

\section{Derivatives of the Lambert $W$ function}\label{appA}

The first derivative of the Lambert $W$ function reads
\begin{equation*}
W'(x)=\frac{W(x)}{x\left[1+W(x)\right]}=\frac{1}{x+e^{W(x)}}
\end{equation*}
and, for $n\geq 1$:
\begin{equation*}
\frac{d^nW(x)}{dx^n}=\frac{e^{-n~W(x)}}{[1+W(x)]^{2n-1}}\mathscr{P}_n(W(x)),
\end{equation*}
where the polynomials $\mathscr{P}_n$ satisfy the recurrence relation \cite{Corless1996}:
\begin{equation*}
\mathscr{P}_{n+1}(y)=-[n(y+3)-1]\mathscr{P}_{n}(y)+(1+y)\mathscr{P}_n'(y),
\end{equation*}
with $\mathscr{P}_1(y)=1$ and $\mathscr{P}_n(0)=(-n)^{n-1}/n!$. The first three polynomials are $\mathscr{P}_1(y)=1$, $\mathscr{P}_2(y)=-2-y$ and $\mathscr{P}_3(y)=9+8y+2y^2$. The general expression of $\mathscr{P}_n(y)$ is
\begin{equation*}
\mathscr{P}_n(y)=(-1)^{n-1}\sum_{k=0}^{n-1}\beta_{n,k}~y^k,
\end{equation*}
where
\begin{equation*}
\beta_{n,k}=\sum_{m=0}^k\frac{1}{m!}\binom{2n-1}{k-m}\sum_{q=0}^m\binom{m}{q}(-1)^q(q+m)^{m+n-1}.
\end{equation*}
Another form is \cite{Dumont1996}:
\begin{equation*}
\frac{d^nW(x)}{dx^n}=
(-1)^{n-1}\left[W'(x)\right]^n\mathscr{Q}_n\left(\frac{1}{1+W(x)}\right),
\end{equation*}
where
\begin{equation*}
\mathscr{Q}_n(y)=\sum_{k=0}^{n-1}b_{n,k}~y^k\;\;\;\;\mathrm{with}\;\;\;\;b_{n,n-1-k}=\sum_{m=k}^{n-1}(-1)^{m-k}\binom{m}{k}\beta_{n,m}.
\end{equation*}
A third possibility consists in using \cite{Knuth2005}:
\begin{equation*}
\frac{d^nW(x)}{dx^n}=\frac{e^{-n~W(x)}}{[1+W(x)]^{n}}\mathscr{R}_n\left(\frac{W(x)}{1+W(x)}\right),
\end{equation*}
where
\begin{equation*}
\mathscr{R}_n(y)=(-1)^{n-1}\sum_{k=0}^{n-1}(-1)^ka_{n,k}~y^k\;\;\;\;\mathrm{with}\;\;\;\;
a_{n,k}=\sum_{m=0}^k(-1)^m\binom{n-1-m}{n-1-k}\beta_{n,m}.
\end{equation*}
One has, for the Lambert function with an exponential argument (which is precisely the case of Eq. (\ref{eq:reci})):
\begin{equation*}
\frac{d^nW(e^x)}{dx^n}=\frac{1}{[1+W(e^x)]^{2n-1}}\mathscr{S}_n(W(e^x)),
\end{equation*}
with
\begin{equation*}
\mathscr{S}_n(y)=\sum_{k=0}^{n-1}\Eulerian{n-1}{k}~(-1)^ky^{k+1},
\end{equation*}
$\Eulerian{a}{b}$ being the Eulerian numbers of second order \cite{Graham1994}. The first derivatives of Lambert's function can also be obtained using
\begin{equation*}
W''(x)=-\frac{f''(W(x))}{[f'(W(x))]^3}
\end{equation*}
and
\begin{equation*}
W'''(x)=3\frac{[f''(W(x))]^2}{[f'(W(x))]^5}-\frac{[f'''(W(x))]}{[f'(W(x))]^4},
\end{equation*}
where $f(y)=y~ e^y$, as a consequence of the Lagrange inversion theorem \cite{Apostol2000,Johnson2002,Jeffrey2015}.

\section{Algebra for the $a~r^k+b~\ln r$ mesh}\label{appB}

\subsection{Implicit case}\label{appB1}

Differentiating
\begin{equation}\label{eq:rklog}
x=a~r^k+b~\ln r
\end{equation}
twice with respect to $r$ gives
\begin{equation*}
\frac{dx}{dr}=\frac{1}{\beta}=\frac{b}{r}+akr^{k-1}, \;\;\;\;\frac{d^2x}{dr^2}=-\frac{b}{r^2}+ak(k-1)r^{k-2}
\end{equation*}
as well as the simple forms
\begin{equation*}
\beta(x)=\frac{r}{b+akr^k}, \;\;\;\;\alpha(x)=\frac{ak(k-1)r^k-b}{(b+akr^k)^2}
\end{equation*}
yielding
\begin{equation*}
\alpha'(x)=\frac{d\alpha}{dx}=\frac{ak^2r^{k-1}\left[b(1+k)-a(k-1)kr^k\right]}{(b+akr^k)^3}\frac{dr}{dx}=\frac{ak^2r^k\left[b(1+k)-a(k-1)kr^k\right]}{(b+akr^k)^4}
\end{equation*}
and
\begin{equation*}
s(x)=\sqrt{\beta(x)}=\sqrt{\frac{r}{b+akr^k}}.
\end{equation*}
Therefore, the differential equation for $t$ is
\begin{equation*}
t''-\frac{1}{(b+akr^k)^2}\left[r^2 q+\frac{b^2+2abk(1+k^2)r^k-a^2k^2(k^2-1)r^{2k}}{4(b+akr^k)^2}\right]t=
\left(\frac{r}{b+akr^k}\right)^{3/2} k
\end{equation*}
with $r \equiv r(x)$ and
\begin{equation*}
t(x)=Y(x)~\sqrt{\frac{b}{r}+akr^{k-1}}.
\end{equation*}
Of course, Eqs. (\ref{eq:loglin}) and (\ref{eq:eqy}) are recovered when $k\rightarrow 1$.

\subsection{Explicit case}\label{appB2}

Inverting Eq. (\ref{eq:rklog}) yields, following the same procedure as in section \ref{subsec:inversion}:
\begin{equation}
r=f(x)=\left(\frac{b}{ak}\right)^{1/k}~\left[W\left(\frac{ka}{b}e^{kx/b}\right)\right]^{1/k}.
\end{equation}
The three first derivatives of $f$ with respect to $x$ read respectively (we set $u=(ka/b)e^{kx/b}$ as intermediate variable):
\begin{equation}
f'(x)=\left(\frac{b}{ak}\right)^{1/k}\frac{[W(u)]^{1/k}}{b}\frac{1}{1+W(u)},
\end{equation}
\begin{equation}
f''(x)=\left(\frac{b}{ak}\right)^{1/k}\frac{[W(u)]^{1/k}}{b^2}\frac{\left[1+(1-k)W(u)\right]}{\left[1+W(u)\right]^3}
\end{equation}
and
\begin{equation}
f'''(x)=\left(\frac{b}{ak}\right)^{1/k}\frac{[W(u)]^{1/k}}{b^3}\frac{\left\{1+\left[2-k(k+3)\right]W(u)+(2k-1)(k-1)W^2(u)\right\}}{\left[1+W(u)\right]^5},
\end{equation}
yielding
\begin{equation}\label{eq:rkun}
- \frac{f'''}{2 f'} + \frac{3(f''){^2}}{4 (f')^2}=\frac{1+2(1+k^2)W(u)-(k^2-1)W^2(u)}{4b^2\left[1+W(u)\right]^4} 
= \frac{b^2+2(1+k^2)abkr^k-(k^2-1)a^2k^2r^{2k}}{4b^2(b+akr^k)^4}, 
\end{equation}
as well as
\begin{equation}\label{eq:rkdeux}
(f')^2 =\frac{1}{b^2}\frac{\left[\displaystyle\frac{b}{ak}W(u)\right]^{2/k}}{\left[1+W(u)\right]^2} 
= \frac{r^2/b^2}{(1+akr^k/b)^2} =\frac{r^2}{(b+akr^k)^2}.
\end{equation}
It is easy to check that Eq. (\ref{eq:rkun}) tends to Eq. (\ref{eq:run}) and Eq. (\ref{eq:rkdeux}) to Eq. (\ref{eq:rdeux}) when $k\rightarrow 1$.


\begin{thebibliography}{99}

\bibitem{Hartree1957} D. R. Hartree, {\it The calculation of atomic structures} (Wiley, New York, 1957).

\bibitem{Numerov1933} B. Numerov, {\it Méthode nouvelle de la détermination des orbites et le calcul des éphémérides en tenant compte des perturbations}, Publ. Observatoire Central Astrophys. Russ. {\bf 2}, 188-288 (1933).

\bibitem{Cooley1961} J. W. Cooley, {\it An improved eigenvalue corrector formula for solving the Schrodinger equation for central fields}, Math. Comp. {\bf 15}, 363-374 (1961).

\bibitem{Hajj1974} F. Y. Hajj, H . Kobeisse and N. R. Nassif, {\it On the numerical solution of Schroedinger’s radial equation}, J. Comput. Phys. {\bf 16}, 150-159 (1974). 

\bibitem{Nikiforov2005} A. F. Nikiforov, V. G. Novikov and V. B. Uvarov, {\it Quantum-Statistical Models of Hot Dense Matter: Methods for Computation Opacity and Equation of State} (Birkh\"auser, 2005).

\bibitem{Havlova1983} H. Havlov\'a and L. Smr$\mathrm{\check{c}}$ka, {\it Numerical solution of the radial Schr\"odinger equation in the inverse cubic scale}, Czech. J. Phys. B {\bf 34}, 961-968 (1984).

\bibitem{Bratcev1965} V. F. Bratcev, {\it Tables of atomic wave functions} (Nauka, Moscow, 1965) [in Russian].

\bibitem{Froese1977} C. Froese-Fischer, {\it The Hartree-Fock method for atoms: a numerical approach} (Wiley-Interscience, 1977).

\bibitem{Pain2006} J.-C. Pain, G. Dejonghe and T. Blenski, {\it A self-consistent model for the study of electronic properties of hot dense plasmas in the superconfiguration approximation}, J. Quant. Spectrosc. Radiat. Transfer {\bf 99}, 451-468 (2006).

\bibitem{Herman1963} F. Herman and S. Skillman, {\it Atomic structure calculations} (Prentice-Hall, Englewood Cliffs, New Jersey, 1963).

\bibitem{Chernysheva1976} L. V. Chernysheva, N. A. Cherepkov and V. Radojevi\'c, {\it Self-consistent field Hartree-Fock program for atoms}, Comput. Phys. Commun. {\bf 11}, 57-73 (1976).

\bibitem{Chernysheva1999} L. V. Chernysheva and V. L. Yakhontov, {\it Two-program package to calculate the ground and excited state wave functions in the Hartree-Fock Dirac approximation}, Comput. Phys. Commun. {\bf 119}, 232-255 (1999).

\bibitem{Wilson2011} B. G. Wilson and V. Sonnad, {\it A note on generalized radial mesh generation for plasma electronic Structure}, High Energy Density Phys. {\bf 7}, 161-162 (2011).

\bibitem{Pain2011} J.-Ch. Pain, {\it Comment on ``A note on generalized radial mesh generation for plasma electronic structure''}, High Energy Density Phys. {\bf 7}, 224 (2011).

\bibitem{Corless1996} R. M. Corless, G. H. Gonnet, D. E. G. Hare, D. J. Jeffrey and D. E. Knuth, {\it On the Lambert W function}, Adv. Comput. Math. {\bf 5}, 329-359 (1996). 

\bibitem{Fernandez2011} F. M. Fern\'andez, {\it Wronskian method for bound states}, Eur. J. Phys. {\bf 32}, 723-732 (2011).

\bibitem{Tannous2008} C. Tannous, K. Fakhreddine and J. Langlois, {\it The Canonical Function Method and its applications in quantum physics}, Phys. Rep. {\bf 467}, 173-204 (2008).

\bibitem{Dumont1996} D. Dumont and A. Ramamonjisoa, {\it Grammaire de Ramanujan et arbres de Cayley}, The Electronic J. Combinatorics {\bf 3}, R17 (1996) [in French].

\bibitem{Knuth2005} D. E. Knuth, {\it The Art of Computer Programming, Vol. 4, Fascicle 3: Generating All Combinations and Partitions, Sec. 7.2.1.5 (Third ed.)} (Reading, Massachusetts: Addison-Wesley,2005).

\bibitem{Graham1994} R. L. Graham, D. E. Knuth and O. Patashnik, {\it Concrete mathematics} (Addison-Wesley, 1994).

\bibitem{Apostol2000} T. M. Apostol, {\it Calculating higher derivatives of inverses}, Amer. Math. Monthly {\bf 107}, 738–741 (2000).

\bibitem{Johnson2002} W. P. Johnson, {\it Combinatorics of higher derivatives of inverses}, Amer. Math. Monthly {\bf 109}, 273-277 (2002).

\bibitem{Jeffrey2015} D. J. Jeffrey, G. A. Kalugin and N. Murdoch. {\it Lagrange Inversion and Lambert W}, In Proceedings of the 2015 17th International Symposium on Symbolic and Numeric Algorithms for Scientific Computing (SYNASC '15). IEEE Computer Society, USA, 42–46 (2015).

\end{thebibliography}
\end{document}